# Problem of Application Job Monitoring in GRID Systems[1]

V. Kalyaev[2], A. Kryukov[3]

Skobektsyn Institute of Nuclear Physics Moscow State University,
119992, Moscow, Russia

We present a new approach to monitoring of the execution process of an application job in the GRID environment. The main point of the approach is use of GRID services to access monitoring information with the security level available in GRID.

## *Introduction.*

The facility of monitoring in GRID [1,2] is very important for effective functionality of the GRID infrastructure. However, the GRID monitoring facilities provide user by the system status of application job. The user can obtain the information whether the job is scheduled, run, canceled or finished. This kind of information is completely enough from GRID management point of view but not enough for users. Let us consider the typical example of a Monte Carlo event simulation for LHC experiments. Usually, a single job generates thousand events for 5-10 hours of CPU. To control the process of generation, a user has to check, from time to time, the current number of generated events. However, there is no any possibility to provide the user with such information in GRID. Therefore, an implementation of the facility for application job monitoring is an important task. In the paper, we propose some approach for solution the problem and describe an experimental realization of the application monitoring in the GRID.

The application monitoring was made in CMS production tools IMPALA/BOSS[3,4,5], McRunJob[5] that are used for MC production on PC clusters (see Fig.1). Omitting some technical details, the IMPALA/BOSS/McRunJob wrap jobs by some script including MySQL client. The script starts a job on a working node of a cluster and parses the standard output to pick up the monitoring information. The extracted information is passed to MySQL client that write it into MySQL database remotely.

[1] This work was partially supported by CERN-INTAS grant 00-0440
[2] kalyaev@theory.sinp.msu.ru
[3] kryukov@theory.sinp.msu.ru

This approach works very well in local case. However there are some restrictions which make difficult its adaptation to the GRID case.

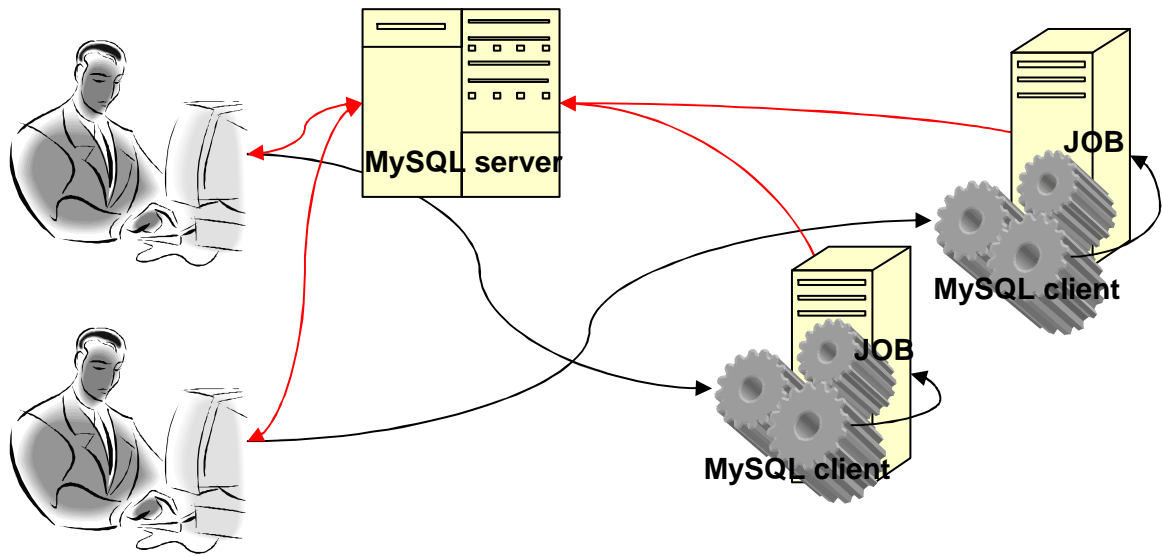

Fig.1. IMPALA/BOSS/McRunJob monitoring scheme.

One of the essential features of a job submission through the GRID is that the user does not know in advance where the job will be executed (See Fig.2.). Thus,
- any submitted jobs have to use the same fixed MySQL server in the world;
- the server has to be opened for write access for the world;
- all jobs have to use the same password for the write access;

It is very difficult to realize this scheme in security way. Moreover, the experience of participating in real MC production shows that ALL users use the same password. So, any user can (potentially) modify and even destroy information on any monitoring server.

Another point is firewall and NAT techniques which are very popular on PC clusters. NAT permits to build a PC cluster where working nodes are not supplied with real IPs. On the other hand, it allows giving them out bounding connection. The availability of out bounding connection is a request of LCG architecture.

The last but not least, it is very difficult to restrict access of different users just to information of their own jobs. This point is not important for production tasks, because all users work in the united team, but very important in the case of real GRID infrastructure where a lot of separate groups work simultaneously.

To solve these problems the new scheme of application job monitoring was proposed.

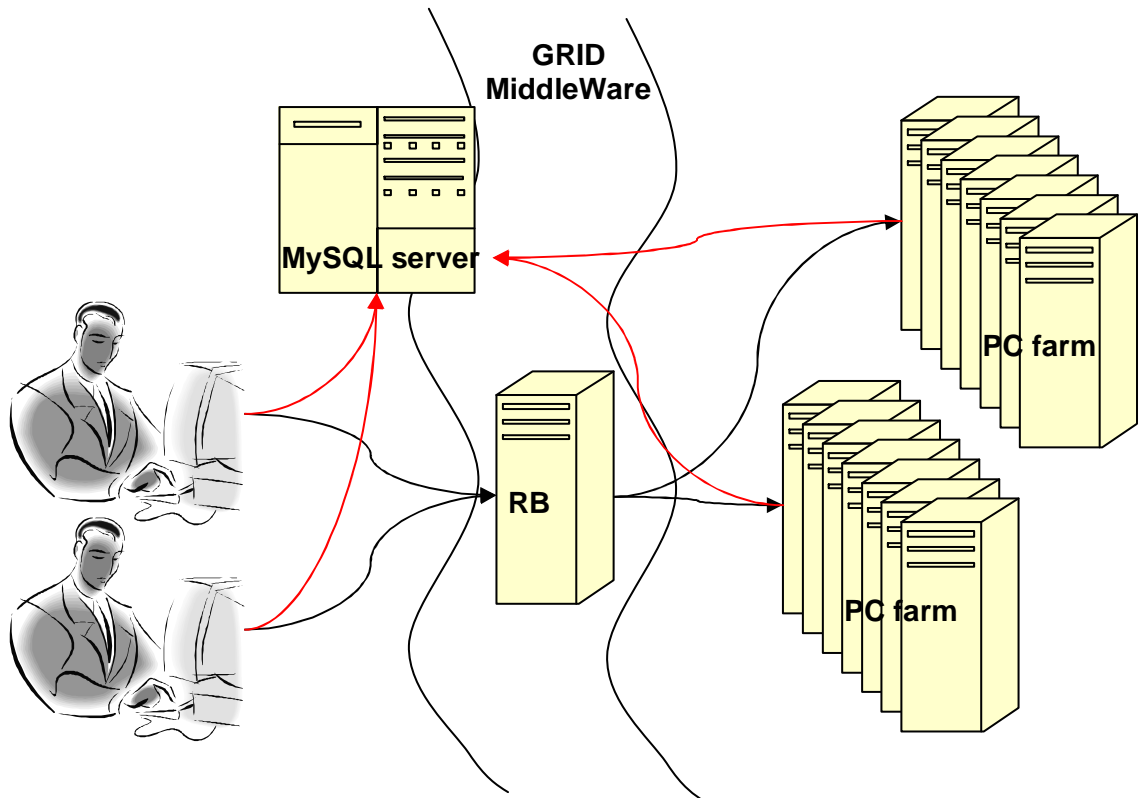

Fig.2. MC event generation in the GRID.

## *New scheme for application job monitoring.*

Let us formulate the main requirements to the application job monitoring system keeping in mind the GRID restrictions:
- secure access to job information by owner of submitted jobs;
- possibility to use many DB servers;
- transparency for end user;
- usage of standard GRID services for realization of the middleware.

To satisfy these requirements, we propose the following scheme of application job monitoring (see Fig.3.).

The main component of proposed scheme is Advance Task Monitor (ATM). This component consists of a set of scripts communicating with MySQL server and providing necessary functionality.

At the first step, a user has to choose the ATM server, which will be used by him for monitoring, and to make a registration. The user starts *atm-user-register* script on User Interface (UI). The script sends the GRID certificate of the user to ATM-server (*atm-user-register-c*) and gets the confirmation or rejection of the registration depending on the site policy. At this step, the default values of user parameters, like maximum number of submitted jobs, are assigned.

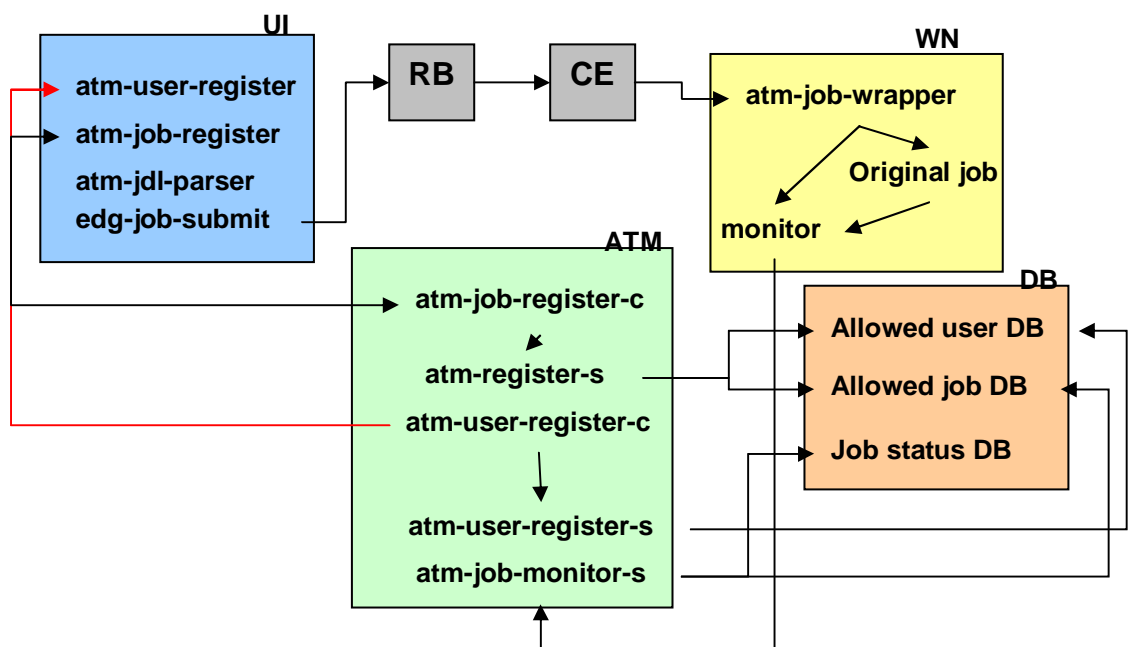

Fig.3. Application job monitoring scheme.

At the second step, the script (*atm-job-register*) is started. This script calls *atm-job-register-c* script on the server side and makes registration of the job. It returns a special ticket for job identification. After that, the job will be wrapped by a special script (*atm-wrapper*) and submitted to a Resource Broker (RB) and further to a Working Node (WN) through some Computing Element (CE).

The wrapper starts monitoring part and original job on WN. From time to time the monitoring part updates monitoring information on the ATM-server by using the job ticket and the user certificate. The script *atm-job-monitor-s* updates the MySQL DB.

All stages use the job ticket and user certificate corresponding to the GSI model of GRID security. So the scheme supports the same level of security as the basic GRID architecture.

There was a change in a JDL file with respect to the new environment of the job (see.

Fig4.). In the modified version of JDL, the parameter **Executable** must be *atm-wrapper*. Therefore the parameter **InputSandbox** must include the file *atm-wrapper*. Also the new parameter **RetryCount** was added. This parameter shows how often the monitoring information will be sent.

The most essential change was made for the parameter **Arguments**. In the modified version it contains some additional information about job like id, password and URL of the executing host (CE). The password is not used for user authorization, but just for the access to the job information.

Of course, all these modifications are automated by the scripts and no special interventions by a user are required.

```
Executable = "atm-wrapper";
StdOutput = "aliroot.out";
StdError = "aliroot.err";
InputSandbox = {"atm-wrapper","start_aliroot2.sh"," rootrc","grun2.C","Confiig.C"};
OutputSandbox = {"aliroot.err","alirot.out","galice.root"};
RetryCount = 10;
Arguments = -id=123 –password=567 –site=test.domain /bin/sh start_aliroot.sh 3.02.04 3.07.01;
Requirements = Member(other.RunTimeEnvironment,"ALICE-3.07.01");
```

Fig.4. Example of JDL files.

## The monitoring information.

In our realization, we have chosen, for simplicity, the standard output as the monitoring information. However, the problem of output delay because of buffering the I/O operation occurs in general case.

Let us consider the case when application output is something like "completed 20 from 200 events" every 10 minutes. The typical buffer size is about 3K. Buffer will be fulfilled for 100*10=1000 minutes, which is for about 16 hours. So in reality, the user does not receive anything at all!

The simplest way to fix the problem is to modify the program code that switches off buffering of the standard output. If this is impossible, the user should consider usage of indirect information like size of resulting file that permit him to understand of application job status[4].

---

[4] Not realized yet.

### *Web interface.*

A Web interface for retrieving by user the necessary information was realized. This is rather straightforward realization based on CGI Perl scripts. However, it removes a lot of duties from the user.

### *Conclusions*

The preliminary tests of the first realization demonstrate that the scheme is working rather well. There are satisfactory scalability and robustness. However, it is necessary to test the middleware in bigger GRID testbed.

In the future, we are going to re-implement the middleware by using the OGSA and to make a comparison of the both approaches.